\documentclass[nofootinbib,onecolumn,showpacs,preprintnumbers,amsmath,amssymb]{revtex4}
\usepackage[dvipdfmx]{graphicx}
\usepackage{amsmath,amssymb}
\usepackage{bm}
\usepackage{color}
\usepackage{hyperref}
\usepackage{enumerate}
\newcommand{\nnmb}{\nonumber\\}
\newcommand{\ket}[1]{\left|{#1}\right>}
\newcommand{\bra}[1]{\left<{#1}\right|}
\newcommand{\innr}[2]{\left<{#1}|{#2}\right>}
\newcommand{\braket}[3]{\left<{#1}\left|{#2}\right|{#3}\right>}

\providecommand{\href}[2]{#2}\begingroup

\usepackage{amsmath}	

\begin{document}

\preprint{YITP-13-47}

\author{Kazuyuki Sugimura$^1$}
\email{sugimura@yukawa.kyoto-u.ac.jp}
\affiliation{%
$^1$Yukawa Institute for Theoretical Physics, Kyoto University, Kyoto, Japan\\
}%

\title{In-in formalism on tunneling background: multi-dimensional quantum mechanics
  }
\begin{abstract}
We reformulate quantum tunneling in a multi-dimensional system where the
tunneling sector is non-linearly coupled to oscillators.  The WKB wave
function is explicitly constructed under the assumption that the system
was in the ground state before tunneling.  We find that the quantum
state after tunneling can be expressed in the language of the
conventional in-in formalism.  Some implications of the result to
cosmology are discussed.
\end{abstract}
  \date{\today}
\maketitle
\section{Introduction}

Quantum tunneling has been studied for a long time as one of the most
exciting topics in various fields of science, from the study of the
dynamics of atomic and molecular systems to condensed matter physics and
field theory (see \cite{mohsen:2003,coleman_aspects}, and references
therein).  Regarding applications to cosmology, there is even a
possibility that the universe was born via quantum
tunneling\cite{Vilenkin:1984wp}.  Furthermore, the string theory
landscape has been proposed as a possible setting of the early universe
inflation\cite{Susskind:2003kw}.  In this framework, scalar fields are
thought to tunnel among many false vacua (i.e. local minima of the
potential) in the vast string theory potential landscape.  The
formulation of the false vacuum decay (i.e. the quantum tunneling from a
false vacuum) in field theory was first considered in flat
spacetimes\cite{Coleman:1977py,Callan:1977pt}, and was extended to
include gravity in \cite{Coleman:1980aw} (see \cite{Sugimura:2011tk} for
the extension to multiple-field cases).

Multi-dimensional quantum tunneling has also been well
studied\cite{mohsen:2003}, and is formulated by constructing the wave
functions for quantum tunneling using the WKB
method\cite{Banks:1973ps,Banks:1974ij,Gervais:1977nv,Yamamoto:1993mp}.
Field theoretic extension was developed in \cite{Tanaka:1993ez},
and such formulation has been applied to the quantum fluctuations on a tunneling
background. 
It was further extended
to include gravity in \cite{Tanaka:1994qa}. As a result of these
developments, it has been possible to calculate the quantum fluctuations
in the universe after false vacuum
decay\cite{Yamamoto:1996qq,Garriga:1997wz,Garriga:1998he}.

All previous works on quantum tunneling neglect effects of non-linear
interactions.  In other words, only free quantum field theory on a
tunneling background has been considered so far.  In light of the recent
progress in observational cosmology, however, it is now important to
study the observational consequences of non-linear interactions.  For
example, the non-Gaussianity of the cosmological fluctuations is now a
hot topic in cosmology
\cite{Bartolo:2004if,Komatsu:2009kd,Chen:2010xka}.  It is clearly
necessary to reformulate quantum field theory on a tunneling background
with non-linear interactions included, in order to calculate the
non-Gaussianity in a universe undergoing quantum tunneling, as is
motivated by the string landscape.  Estimates for the non-Gaussianity
in such a scenario have been calculated 
in the literature\cite{Sugimura:2012kr,Park:2011ty}, but up to now
there is no rigorous proof that the formulation used there is valid.

In this paper, we reformulate multi-dimensional quantum tunneling
with non-linear interactions, following the formulation by Yamamoto
\cite{Yamamoto:1993mp}. Although the formulation of the
multi-dimensional system is interesting in itself, it can also be
regarded as a first step towards the formulation of quantum field
theory. 
We expect that extensions from multi-dimensional cases to 
field theory with gravitation are possible as before
\cite{Yamamoto:1993mp,Tanaka:1993ez,Tanaka:1994qa},
but leave such issues to future studies.

\begin{figure}
\includegraphics[width=12cm]{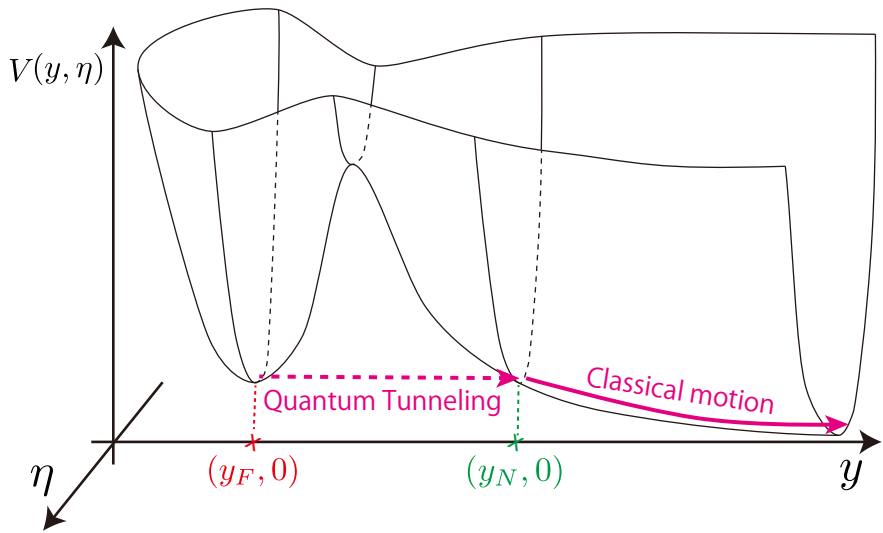}  
\caption{The potential for the 2-dimensional system, where the tunneling sector $y$ is non-linearly coupled to
the oscillator $\eta$.
The particle moves from the false vacuum $(y_F, 0)$ 
to the nucleation point $(y_N,0)$ by quantum tunneling,
and rolls down classically from the nucleation point.
}
\label{fig:potential}
\end{figure}

As the simplest extension of the 1-dimensional case, we will study a
2-dimensional system in which the tunneling sector $y$ is non-linearly
coupled to the oscillator $\eta$, as shown in Fig.~\ref{fig:potential}.
The restriction to a 2-dimensional system keeps calculations as simple
as possible whilst still maintaining the essential features of
multi-dimensional effects.  The particle, originally positioned in the
false vacuum at $(y_F,0)$, moves to the nucleation point at $(y_N,0)$ by
quantum tunneling, and then rolls down classically, as shown in
Fig.~\ref{fig:potential}.  Assuming that the potential is static, the
wave function $\Psi(y,\eta)$ for such a particle is a solution of the
time-independent Schr\"odinger equation. The boundary conditions for
$\Psi(y,\eta)$ corresponding to the scenario outlined above are given as
follows: $\Psi(y,\eta)$ should be an out-going wave function outside the
barrier, and $\Psi(y,\eta)$ should match the wave function for the
quantum state before the quantum tunneling around the false vacuum.

Let us put a screen at $y$ outside the barrier,
and then prepare the above system many times and let the particles hit the screen.
The particles hit the screen with
different $\eta$ each time, since $\Psi(y,\eta)$ is extended
in the $\eta$ direction.
The statistical properties of $\eta$ at $y$
are given by the quantum expectation values with respect to $\Psi(y,\eta)$, defined as
$\big<\eta^n\big>_{y}\equiv \int d\eta \eta^n|\Psi(y,\eta)|^2$ where $n=1,\,2,\,3,\,\cdots$.
In this paper, we obtain formulae for such quantities
by constructing $\Psi(y,\eta)$ explicitly using the WKB method.
If we define $t$ as the time the particle takes to reach $y$ from the nucleation point,
we can interpret $\Psi(y,\eta)$ as the $t$-dependent wave function with respect to $\eta$.
Then, we find that our resulting formulae can be expressed
in the language of the conventional in-in formalism\cite{Weinberg:2005vy,Maldacena:2002vr}.
Note that
$\big<\eta^n\big>_y$ at given $y$, or given $t$, can be regarded as the analogue
of the $n$-point correlation functions at a given time in field theory,
where the time is defined in terms of the value of the tunneling field.

This paper is organized as follows.  In Sec.~\ref{sec:wkbnext}, we
obtain the expression for the quantum expectation value in the
Schr\"odinger picture.  In Sec.~\ref{sec:interaction}, we move to the
interaction picture, where the quantum expectation value is given in the
in-in formalism form.  In Sec.~\ref{sec:appl-toy-model}, we apply the
formalism obtained in Sec.~\ref{sec:wkbnext} and in
Sec.~\ref{sec:interaction} to a simple toy model for illustration
purposes.  Finally, we conclude in Sec.~\ref{sec:conclusion}.

\section{Formulation: Schr\"odinger picture}
\label{sec:wkbnext}

\subsection{WKB analysis for 2-dimensional system}
As mentioned in the introduction,
let us consider a 2-dimensional system.
The Hamiltonian of the system is given by
\begin{align}
\mathcal{H}&=\frac{p_y^2}{2}+\frac{p_\eta^2}{2}+V(y,\eta)\,,
\label{eq:8}
\end{align}
where $V(y,\eta) $ has a false vacuum and nucleation point at
$(y,\eta)=(y_F,0)$ and $(y_N,0)$, respectively, as shown in
Fig.~\ref{fig:potential}. The nucleation point is defined as the
opposite end to the false vacuum on the tunneling path, which is the
classical trajectory connecting the false vacuum and the region outside
the potential barrier with minimum action.  Separating $V(y,\eta)$
into the $y$-part $V_{tun}(y)$ and the $\eta$-part $V_\eta(y,\eta)$ as
$V(y,\eta)=V_{tun}(y)+V_\eta(y,\eta)$, we assume for simplicity that
$V_\eta(y,\eta)$ can be written as
$V_\eta(y,\eta)=(\omega^2(y)/2)\eta^2+V_{int}(y,\eta)$, where
the nonlinear interaction term $V_{int}(y,\eta)$ consists of 
the cubic and higher order terms with respect to $\eta$.
The vanishing of the linear term with respect to $\eta$ in the potential
guarantees that the tunneling path lies on the $y$-axis.
The inclusion of the nonlinear interaction term 
$V_{int}(y,\eta)$ is the essential new point in this paper,
compared to the literature\cite{Yamamoto:1996qq,Garriga:1997wz,Garriga:1998he}.
For later convenience, here we denote
the $y$- and $\eta$-parts of the Hamiltonian as
$\mathcal{H}_y=p_y^2/2+V_{tun}(y)$ and $\mathcal{H}_\eta =
p_\eta^2/2+V_\eta(y,\eta)$, respectively.

In the system defined by eq.~\eqref{eq:8}, we consider 
the tunneling wave function $\Psi(y,\eta)$, which is a solution of the
time independent Schr\"odinger equation with eigenenergy $E$
\begin{align}
 \hat{\mathcal{H}}\Psi(y,\eta)&= E\Psi(y,\eta)\,.
\label{eq:10}
\end{align}
Here, quantities with hat($\hat{\ }$) are operators, and $\hat{p}_y$ and
$\hat{p}_\eta$ in $\hat{\mathcal{H}}$ are given by
$(\hbar/i)(\partial/\partial y)$ and $(\hbar/i)(\partial/\partial
\eta)$, respectively.  
In this paper, we concentrate on quantum
tunneling from the quasi-ground-state, which is defined as the ground state 
for the potential expanded around the false vacuum.
We can consider quantum tunneling from excited states,
as in \cite{Yamamoto:1993mp}, but we leave such issues to future studies.
As mentioned in the introduction, 
$\Psi(y,\eta)$ should be an out-going wave function outside the barrier.

We construct the tunneling wave function under the following assumptions:
\begin{enumerate}[\hspace{1cm}1)]
\item the WKB approximation is valid well inside and well outside the barrier,
\item the coupling between the $y$ and $\eta$ directions is small,
\item the region around the nucleation point where the WKB approximation breaks is narrow,
\item the coupling between the $y$ and $\eta$ directions vanishes around the false vacuum.
\end{enumerate}
We hope to return to more general cases, say, cases where assumptions 3)
and/or 4) are relaxed, in future.
If there was no coupling between the two
directions (i.e. if $V_\eta(y,\eta)$ could be denoted as $V_\eta(\eta)$), the
tunneling wave function $\Psi(y,\eta)$ would be given by the product of
$\Psi_y(y)$ and $\Phi(\eta)$, where $\Psi_y(y)$ is the
1-dimensional tunneling wave function for $V_{tun}(y)$ and
$\Phi(\eta)$ is the ground state for $V_\eta(\eta)$. 
In our case, however, we consider small but non-vanishing coupling,
and thus we expand $\Psi(y,\eta)$ and $E$ in eq.~\eqref{eq:10} as
\begin{align}
 \Psi(y,\eta)=\Psi_y(y)\Phi(y,\eta)\,,\qquad
E=E_y+E_\eta\,.
\label{eq:11}
\end{align}
Here, $\Psi_y(y)$ and $E_y$ are, respectively, the wave function and energy of
the 1-dimensional Schr\"odinger equation $\mathcal{H}_y\Psi_y(y)=E_y\Psi_y(y)$,
which we will briefly discuss below.
As a result of assumption 4),
the quasi-ground-state is given by $\Psi_y(y)\Phi_F(\eta)$,
where $\Phi_F(\eta)$ is the ground state for
the $\eta$-part of the potential around the false vacuum $V_F(\eta)(\equiv V_\eta(y_F,\eta))$. 
Here, by focusing on eq.~\eqref{eq:10} around the false vacuum
and denoting the ground state energy with respect to $V_F(\eta)$ as $E_F$,
it can be seen that $E_\eta$ is given by $E_F$.

\begin{figure}
 \includegraphics[width=9cm]{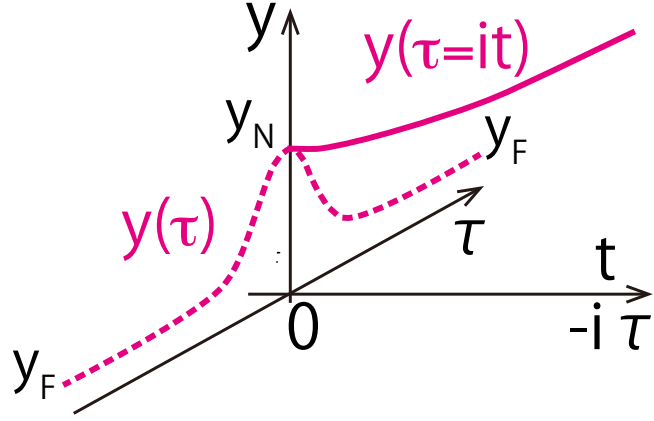} \caption{The 
schematic picture of instanton $y(\tau)$ with the imaginary time $\tau$
(doted line)
and its analytical continuation $y(t)\equiv y(\tau=it)$
with the real time $t$ (solid line).} \label{fig:inst}
\end{figure}

As shown in Fig.~\ref{fig:inst}, the tunneling path $y(\tau)$, or instanton, is a
solution of the Euclidean equation of motion (EOM)
$y''(\tau)-dV_{tun}/dy=0$, where $'$ denotes the derivative with respect
to the imaginary, or Euclidean, time $\tau$. The boundary conditions for $y(\tau)$ are given by
$y(\pm \infty)= y_F$ and $y(0)=y_N$, where
the freedom in choosing the origin of $\tau$ is fixed. 
Well inside the potential barrier,
we rewrite the wave function as $\Psi_y(y)=e^{-S_y(y)/\hbar}$ 
with the Euclidean action $S_y(\tau)(=S_y(y(\tau)))$,
and make the WKB expansion $S_y=S_0+\hbar S_1+ \hbar^2 S_2+\cdots$.
Then, by solving the Schr\"odinger equation order by order
and using the instanton $y(\tau)$, we
can obtain $dS_0(y)/dy= y'(\tau)$, $S_1(y)=(1/2)\ln (dS_0/dy)$, and so on,
where we take $\tau$ to be in the region $\tau\in (-\infty,0)$.
It is known that we can move from inside the barrier to outside the barrier
by analytical continuation $\tau\to t=-i\tau$, where $t$ is the real, or Lorentzian, time.
After the analytical continuation, the instanton gives
the classical motion of the particle
$y(t)\equiv y(\tau=it)$,
which starts rolling down from the nucleation point
at $t=0$, as shown in Fig.~\ref{fig:potential} and Fig.~\ref{fig:inst}.
Furthermore, the analytical continuation of the Euclidean action $S_y(t)\equiv S_y(\tau=it)$
gives the tunneling wave function $\Psi_y(y(t))=e^{-S_y(t)/\hbar}$
well outside the barrier.
In the following, we can use $\tau$, $t$ and $y$ interchangeably.

Now, we will transform eq.~\eqref{eq:10} inside the potential barrier.  By
substituting eq.~\eqref{eq:11} with $E_\eta=E_F$ into eq.~\eqref{eq:10} and using
the 1-dimensional Schr\"odinger equation  $\mathcal{H}_y\Psi_y(y)=E_y\Psi_y(y)$,
we obtain
\begin{align}
 \hbar\frac{dS_y}{dy}\frac{\partial }{\partial y}\Phi(y,\eta)
-\frac{\hbar^2}{2}\frac{\partial^2 }{\partial y^2}\Phi(y,\eta)
+\hat{H}(y)\Phi(y,\eta)=0\,,
\label{eq:2}
\end{align}
where 
\begin{align}
 \hat{H}(y)&=\frac{\hat{p}_\eta^2}{2}+V_\eta(y,\eta)-E_F\,.
\label{eq:60}
\end{align}
Here, we can neglect the second term in eq.~\eqref{eq:2},
since the $y$-dependence of $\psi(y,\eta)$
is expected to be small as a result of assumption 2).  
By neglecting the
second term in eq.~\eqref{eq:2} and using the leading order relation in
the WKB approximation $\hbar(dS_y/dy)(\partial/\partial y)\approx
\hbar(\partial/\partial \tau)$, we can transform eq.~\eqref{eq:2} into
\begin{align}
-\hbar\frac{\partial}{\partial\tau}\Phi(\tau,\eta)
=\hat{H}(\tau)\Phi(\tau,\eta)\,.
\label{eq:29}
\end{align}
This equation
is of exactly the same form as the ``time-dependent Schr\"odinger equation''
with imaginary time $\tau$, defined for $\tau\in(-\infty,0)$.

Let us now check the consistency of the approximation 
used to derive eq.~\eqref{eq:29},
by estimating the size of the second term in eq.~\eqref{eq:2}.
To next-to-leading order in the WKB approximation,
the coefficient of the second term in eq.~\eqref{eq:2} can be approximated as 
\begin{align}
\frac{\hbar^2}{2}\frac{\partial^2 }{\partial y^2}&\approx
 -\frac{\hbar y''}{{y'}^3}\hbar\frac{\partial }{\partial \tau}
 +\frac{1}{2{y'}^2}(\hbar\frac{\partial}{\partial \tau})^2\,.
\label{eq:58}
\end{align}
Here, ${\hbar y''}/{{y'}^3}\approx 2(dS_1(y)/dy)/(dS_0(y)/dy)$
and $(\hbar{\partial }/{\partial \tau})$ can be estimated 
as $O(\hbar \omega)$ using eq.~\eqref{eq:29}.
Thus, when the first and second operators on the r.h.s. act
on $\Phi(\tau,\eta)$,
they give terms that are suppressed, under WKB approximation, by
factors of $O((dS_1(y)/dy)/(dS_0(y)/dy))$ and $O(\hbar \omega/y'^2)$
relative to other terms in eq.~\eqref{eq:2}, respectively.

It may be useful to make a comment on the WKB expansion used above.
Strictly speaking, this expansion is not merely a expansion in  $\hbar$
where $\eta$ is considered to be $O(\hbar^{1/2})$, as was done in \cite{Yamamoto:1993mp}. 
In such an expansion,
the non-linear interaction terms would not appear in eq.~\eqref{eq:29},
since the non-linear interaction terms would become higher order in $\hbar$
(e.g. $\eta^3$ term would become $O(\hbar^{3/2})$).
Rather, here we have expanded equations based on the fact that the classical part
of the wave function $S_0(y)$ dominates over quantum effects,
which makes it possible to consistently take into account the effect of non-linear interaction terms
in eq.~\eqref{eq:29}.

We can also transform eq.~\eqref{eq:10} outside the barrier, following similar arguments to those
outlined above
but with the real time $t$ instead of the imaginary time $\tau$.
As a result of the analytical continuation $\tau\to t=-i\tau$, we obtain 
\begin{align}
i\hbar\frac{\partial}{\partial t}\Phi(t,\eta)
=\hat{H}(t)\Phi(t,\eta)\,,
\label{eq:3}
\end{align}
which is the ``time-dependent Schr\"odinger equation'' with real time $t$, defined for $t\in(0,\infty)$.
For later convenience, let us recall that the original 2-dimensional wave function $\Psi(y,\eta)$ 
is denoted as
\begin{align}
 \Psi(y,\eta)&=\exp\left[{-{S(t)}/{\hbar}}\right]\Phi(t,\eta)\,,
\label{eq:13}
\end{align} 
where $y(=y(\tau=it))$ is inside and outside the potential barrier
for $t\in(+i \infty,0)$ and $t\in(0,\infty)$, respectively.

Around the false vacuum or the nucleation point,
where the WKB approximation is not valid,
we determine $\Phi$ using matching conditions.
Thanks to assumptions 3) and 4),
the matching conditions are given in a simple way.
Firstly, the matching condition at $y=y_N$ is given by 
\begin{align}
\lim_{\tau\to -0}\Phi(\tau,\eta)=\lim_{t\to +0}\Phi(t,\eta)\,,
\label{eq:61}
\end{align}
since $\Psi(y,\eta)=\Psi_y(y)\Phi(y,\eta)$ on both sides of $y_N$
should have the same value at $y_N$. Here, we can use eq.~\eqref{eq:29}
and eq.~\eqref{eq:3} until very close to $y_N$
thanks to assumption 3).  Secondly, the matching condition at $y=y_F$ is
given by
\begin{align}
 \lim_{\tau\to-\infty }\Phi(\tau,\eta)= \Phi_F(\eta)\,,
\label{eq:62}
\end{align}
since the wave function is assumed to match
the quasi-ground-state around the false vacuum,
which is given by $\Psi_y(y)\Phi_F(\eta)$ due to assumption 4),
as mentioned below eq.~\eqref{eq:11}.

\subsection{Expectation values of operators}
We will obtain the tunneling wave function
by solving eq.~\eqref{eq:29} and eq.~\eqref{eq:3} with the matching condition
eq.~\eqref{eq:61} and eq.~\eqref{eq:62}.
For notational simplicity, we introduce bra-ket notation,
where eq.~\eqref{eq:3} is written as
\begin{align}
 i\hbar\frac{\partial}{\partial t}\ket{\Phi(t)}&=\hat{H}(t)\ket{\Phi(t)}\,,
\label{eq:12}
\end{align}
with
\begin{align}
\innr{\eta}{\Phi(t)}&=\Phi(t,\eta)\,.
\end{align}
The formal solution to eq.~\eqref{eq:12} is given by
\begin{align}
\ket{\Phi(t)}=
P\left(\exp\left[-\frac{i}{\hbar}\int_{t_0}^t H(t')dt'\right]\right)\ket{\Phi(t_0)}\,,
\label{eq:6}
\end{align}
where $0<t_0<t$ and
the path ordering operator $P$ orders operators
according to their order along the integration path.
From now on, we omit $\hat{\ }$ over operators for brevity.
Similarly, the formal solution to eq.~\eqref{eq:29} 
is given by
\begin{align}
\ket{\Phi(\tau)}=
P\left(\exp\left[-\frac{1}{\hbar}\int_{-i\tau_0}^{-i\tau} H(\tau')d\tau'\right]\right)\ket{\Phi(\tau_0)}\,,
\label{eq:7}
\end{align}
for $\tau_0<\tau<0$.
The expressions in eq.~\eqref{eq:6} and eq.~\eqref{eq:7} are not valid at the nucleation point,
where the WKB approximation breaks down.
However, thanks to the matching condition given by
eq.~\eqref{eq:61}, which can be written in bra-ket notation
as $\ket{\Phi(\tau=-0)}=\ket{\Phi(t=+0)}$,
we can connect the two expressions at the nucleation point as
\begin{align}
\ket{\Phi(t)}
&=P\left(\exp\left[-\frac{i}{\hbar}\int_{0}^{t} H(t')dt'\right]\right)\ket{\Phi(0)} \nnmb
&=P\left(\exp\left[-\frac{i}{\hbar}
\int_{-i\tau_0\to0\to t} H(t')dt'\right]\right)\ket{\Phi(-i\tau_0)}\,,
\label{eq:4}
\end{align}
where $\int_{-i\tau_0\to0\to t}dt'=\int_{0}^{t}dt'+\int_{-i\tau_0}^{0}dt'$.

The matching around the false vacuum is given as follows. We consider a wave function
which matches the quasi-ground-state around the false vacuum.
The ket $\ket{\Omega_F}$ corresponding to the quasi-ground state $\Phi_F(\eta)$
can be given by
 \begin{align}
 \ket{\Omega_F}&=
\lim_{T\to\infty}e^{-\frac{1}{\hbar}H_FT}\ket{\Phi}\,,
 \label{eq:5}
 \end{align}
where $H_F\equiv H(+i\infty)$ and 
$\ket{\Phi}$ is arbitrary as long as it is not orthogonal to $\ket{\Omega_F}$.
We don't need to care about the overall normalization of  $\ket{\Omega_F}$,
since it will be canceled in the calculations of
quantum expectation values, as will be seen below.
In deriving eq.~\eqref{eq:5}, we use the fact that the ground state has $H_F=0$
while excited states have $H_F>0$, which comes from the definition of
$H(y)$ in eq.~\eqref{eq:60}.
From assumption 4), there exists a $\tau_0$ such that for $\tau<\tau_0$
we can approximate $H(\tau)$ and $\ket{\Phi(-i\tau)}$
as $H_F$ and $\ket{\Omega_F}$, respectively.
Thus, using eq.~\eqref{eq:4} and eq.~\eqref{eq:5},
the state evolving from $\ket{\Omega_F}$ at $t=-i\tau_0$ is given by
\begin{align}
 \ket{\Phi(t)}
&=
P\left(\exp\left[-\frac{i}{\hbar}\int_{-i\tau_0 \to 0\to t} H(t')dt'\right]\right)
\lim_{T\to\infty}e^{-\frac{1}{\hbar}H_FT}\ket{\Phi}\nnmb
&=P\left(\exp\left[-\frac{i}{\hbar}\int_{+i\infty\to0\to t} H(t')dt'\right]\right)\ket{\Phi}\,.
\label{eq:15}
\end{align}

\begin{figure}
\includegraphics[width=9cm]{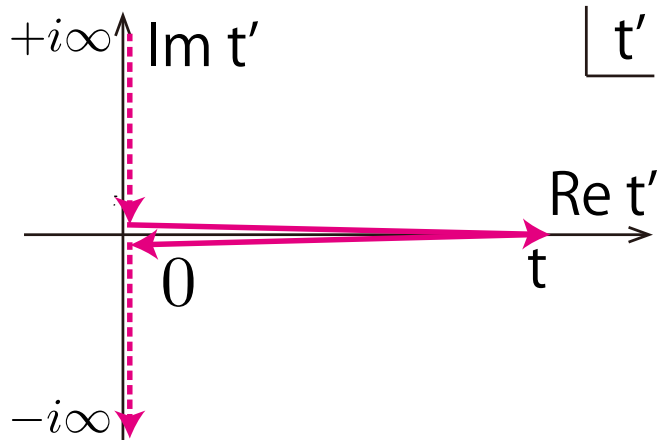} \caption{The time
integration path $C$ given by eq.~\eqref{eq:32}.
  The time integration along the imaginary axis (doted
line) corresponds to the evolution of the quantum state during
tunneling, and along the real axis (solid line) corresponds to the
evolution after tunneling.  } \label{fig:inin}
\end{figure}

Now we are able to evaluate the quantum expectation values.  For an
operator $\mathcal{O}$ with respect to $\eta$ (i.e. some function of
$\eta$ and $p_\eta$), the quantum expectation value at given $y(=y(t))$
outside the barrier is given by
\begin{align}
 \big<\mathcal{O}\big>_y&=
\frac{\int_{-\infty}^\infty d\eta \Psi^*(y,\eta)\mathcal{O}\Psi(y,\eta)}
{\int_{-\infty}^\infty d\eta |\Psi(y,\eta)|^2}\nnmb
&=
\frac{\braket{\Phi(t)}{\mathcal{O}}{\Phi(t)}}
{\innr{\Phi(t)}{\Phi(t)}}\,.
\label{eq:17}
\end{align}
To derive the second line,
we use eq.~\eqref{eq:13} and cancel the factors $e^{-\frac{1}{\hbar}S(t)}$
appearing both in numerator and denominator.
Taking the hermitian conjugate of eq.~\eqref{eq:15}, we obtain
\begin{align}
 \bra{\Phi(t)}&=\left(\ket{\Phi(t)}\right)^\dagger\nnmb
&=
\bra{\Phi}P\left(\exp\left[-\frac{i}{\hbar}
\int_{t\to0\to -i\infty} H(t')dt'\right]\right)\,,
\label{eq:19}
\end{align}
where $H(t^*)=H(t)$ since $H(y)$ given in eq.~\eqref{eq:60} depends only on $y$
and $y(-\tau)=y(\tau)$ due to the Euclidean time inversion symmetry of the instanton. 
By substituting eq.~\eqref{eq:15} and eq.~\eqref{eq:19} into eq.~\eqref{eq:17},
we obtain the resulting formula for the quantum expectation
values in the Schr\"odinger picture
\begin{align}
\big<\mathcal{O}\big>_y&=
\frac{\braket{\Phi}{
P\left(\mathcal{O}\exp\left[-\frac{i}{\hbar}
\int_C
H(t')dt'\right]
\right)
}{\Phi}}
{\braket{\Phi}{P\left(\exp\left[-\frac{i}{\hbar}
\int_C H(t')dt'\right]\right)}{\Phi}}\,,
\label{eq:16}
\end{align}
where 
\begin{align}
 C:+i\infty\to0\to t\to0\to-i\infty
\label{eq:32}
\end{align}
is the time integration path,
as shown in Fig.~\ref{fig:inin},
and $\mathcal{O}$ is ordered by $P$ as if it is defined at $t$.
In the denominator of eq.~\eqref{eq:16}, we can deform the integration path
from $C$ to $i\infty\to -i\infty$
using $P\left(\exp\left[-\frac{i}{\hbar}\int_{0\to t\to 0}
H(t')dt'\right]\right)=1$.
If $\ket{\Phi}$ was chosen to be orthogonal to $\ket{\Omega_F}$, we could obtain the quantum expectation values
for quantum tunneling from an excited state, as studied in
\cite{Yamamoto:1993mp}. We leave such issues to future studies.

\section{Formulation: interaction picture}
\label{sec:interaction}

\subsection{Relation between interaction and Schr\"odinger pictures}
Since the expression given in eq.~\eqref{eq:16} is difficult to evaluate
directly, in this section we will move from the Schr\"odinger
picture formulation to the interaction picture
one. 
This can be accomplished almost in the same way as usual,
but taking into account the non-unitarity of the evolution operator
for the imaginary part of the integration path.
The interaction picture formulation may be helpful when considering the 
multi-dimensional tunneling system in the context of
quantum field theory, where the interaction picture is employed.

First of all, we introduce the evolution operator
\begin{align}
 U(t_2,t_1)&=P\left(\exp\left[-\frac{i}{\hbar}\int_{t_1}^{t_2} H(t)dt\right]\right)\nnmb
&\equiv 1+(-i)\int_{t_1}^{t_2}H(t')dt'
+(-i)^2\int_{t_1}^{t_2}dt'\int_{t_1}^{t'}dt''H(t')H(t'')+\cdots\,,
\label{eq:27}
\end{align}
where $t_1$ and $t_2$ are on the path $C$ given by eq.~\eqref{eq:32}.
The inverse operator for $U(t_2,t_1)$ is given by
\begin{align}
 \left(U(t_2,t_1)\right)^{-1}&=U(t_1,t_2)\,,
\label{eq:33}
\end{align}
which can be confirmed by explicit calculation of
$(U(t_2,t_1)\left(U(t_2,t_1)\right)^{-1}$ using eq.~\eqref{eq:27}.
The combination rule 
\begin{align}
 U(t_3,t_2)U(t_2,t_1)&=U(t_3,t_1)\,,
\end{align}
is satisfied as usual.
It should be noted that $U(t_2,t_1)$ is not generally a unitary operator
since the path $C$ include the imaginary part,
and that $U(t_2,t_1)$ satisfies the relation $U(t_2,t_1)^{\dagger}=U(t_2^*,t_1^*)^{-1}$.

To find interaction picture expression,
we expand the full Hamiltonian given in eq.~\eqref{eq:60} as $H(t)=H_0(t)+H_{int}(t)$,
where the free part $H_0(t)$ and the interaction part $H_{int}(t)$ 
are given, respectively, by
\begin{align}
 H_{0}(t)=\frac{p_\eta^2}{2}+\frac{\omega^2(t)}{2}\eta^2-E_F\,,\qquad
H_{int}(t)=V_{int}(y(t),\eta)\,.
\label{eq:47}
\end{align}
Using $H_{0}(t)$,
we can define the annihilation and creation operators at each $t$, respectively, as
\begin{align}
 a_{t}&=\sqrt{\frac{2\omega(t)}{\hbar}}\eta+i\sqrt{\frac{2}{\hbar\omega(t)}}p_\eta\,,\qquad 
a^\dagger_{t}=\sqrt{\frac{2\omega(t)}{\hbar}}\eta-i\sqrt{\frac{2}{\hbar\omega(t)}}p_\eta\,,
\label{eq:59}
\end{align}
where $a_{t}$ and $a^\dagger_{t}$ satisfy the usual commutation relation.
The eigenstates with respect to $H_0(t)$ can be defined with $a_{t}$ and
$a^\dagger_{t}$ as
\begin{align}
 \ket{n_t}=\frac{1}{\sqrt{n!}}\left(a^\dagger_{t}\right)^n\ket{0_t}\,,\qquad
a_{t}\ket{0_t}=0\,,
\label{eq:18}
\end{align}
where they satisfy
\begin{align}
 H_{0}(t)\ket{n_t}&=E_{n_t}\ket{n_t}\,,\qquad
E_{n_t}=\hbar\omega(t)\left(n+\frac{1}{2}\right)-E_F\,.
\label{eq:34}
\end{align}
When $H_0(t)$ explicitly depends on time,
$a_t$ and $\ket{0_t}$ also become time-dependent.
$a_t$ at the times $t=t_1$ and $t=t_2$ are related by a 
Bogolubov transformation,
and $\ket{0_{t_1}}$ and $\ket{0_{t_2}}$
are annihilated by $a_{t_1}$ and $a_{t_2}$, respectively.
The evolution operator for the free Hamiltonian $H_0(t)$
is given by
\begin{align}
  U^{(0)}(t_2,t_1)&=
P\left(\exp\left[-\frac{i}{\hbar}\int_{t_1}^{t_2} H_0(t)dt\right]\right)\,.
\label{eq:45}
\end{align}

Interaction picture operators $\mathcal{O}_I(t)$ are defined by
\begin{align}
 \mathcal{O}_I(t)&\equiv U^{(0)}(0,t)
\mathcal{O}\, U^{(0)}(t,0)\,,
\label{eq:35}
\end{align}
where $\mathcal{O}$ are Schr\"odinger picture operators.
In the interaction picture, states 
are evolved with the evolution operator for $H_I(t)$, given by
\begin{align}
  U_I(t_2,t_1)&=
P\left(\exp\left[-\frac{i}{\hbar}\int_{t_1}^{t_2} H_I(t)dt\right]\right)\,,
\label{eq:21}
\end{align}
where the interaction Hamiltonian $H_I(t)$ is defined as
\begin{align}
 H_I(t)&\equiv H_{int}(\eta_I(t),t)\,.
\label{eq:40}
\end{align}
For any $t_1$ and $t_2$ on $C$ given by eq.~\eqref{eq:32},
we can rewire $U_I(t_2,t_1)$ in terms of $U(t_2,t_1)$ from eq.~\eqref{eq:27}
and $U^{(0)}(t_2,t_1)$ from eq.~\eqref{eq:45} as
\begin{align}
  U_I(t_2,t_1)&=U(t_2,t_1) U^{(0)}(t_1,t_2)
 = U^{(0)}(t_1,t_2)U(t_2,t_1)\,,
\label{eq:39}
\end{align}
which can be confirmed by explicit calculation.

To describe $\eta_I(t)$ and $p_{\eta I}(t)$ in a simple way, 
we introduce a positive frequency function $u(t)$
and a negative frequency function $v(t)$.
They are defined as solutions to the linearized EOM,
\begin{align}
 \ddot{u}(t)=-\omega^2(t)u(t)\,,\qquad \ddot{v}(t)=-\omega^2(t)v(t)\,,
\label{eq:38}
\end{align}
which are complex conjugate to each other when $t$ is real;
\begin{align}
 u^*(t)=v(t)\qquad{\rm for\ real\ } t\,,\label{eq:14}
\end{align}
and satisfy Klein-Goldon(KG) normalization,
\begin{align}
u(t)\dot{v}(t)-\dot{u}(t)v(t)=i\hbar\,.
\label{eq:23}
\end{align}
Here, a dot denotes the derivative with respect to $t$. When $t$ is
imaginary, since we define $u(t)$ and $v(t)$ by analytical continuation from
real $t$, eqs.~\eqref{eq:38} and \eqref{eq:23} still hold
but eq.~\eqref{eq:14} is no longer true.  It should be noted that the freedom in
choosing $u(t)$ corresponds to the freedom to make an arbitrary Bogolubov
transformation.

Using $u(t)$ and $v(t)$, we can define the annihilation operator $a$ and
the creation operator $a^\dagger$, respectively,  as
\begin{align}
a=-\frac{i}{\hbar}\left(\eta_I(t)\dot{v}(t)-p_{\eta I}(t)v(t)\right)\,,\qquad
a^\dagger=\frac{i}{\hbar}\left(\eta_I(t)\dot{u}(t)-p_{\eta I}(t)u(t)\right)\,.
\label{eq:25}
\end{align}
We will see below that the operators defined in eq.~\eqref{eq:25}
are time-independent and hermitian conjugate to each other.  Firstly, it can be
explicitly shown that these operators are time-independent by 
differentiating $a$ and $a^\dagger$ in eq.~\eqref{eq:25} with respect to $t$
and using eq.~\eqref{eq:38} and the evolution equations for $\eta_I(t)$ and
$p_{\eta I}(t)$,
\begin{align}
 \dot{\eta}_I(t)&=\frac{1}{i\hbar}\left[\eta_I(t),H_0(t)\right]
=p_{\eta I}(t)\,,\qquad
 \dot{p}_{\eta I}(t)=\frac{1}{i\hbar}\left[p_{\eta I}(t),H_0(t)\right]
=-\omega^2(t)\eta_ I(t)\,.
\label{eq:24}
\end{align}
Since eqs.~\eqref{eq:38} and \eqref{eq:24} are valid not only for
real $t$ but also for imaginary $t$, eq.~\eqref{eq:25} can be used even
when $t$ is imaginary. Secondly, by considering eq.~\eqref{eq:25} when
$t$ is real and using eq.~\eqref{eq:14} and the hermiticity
of $\eta_I(t)$ and $p_{\eta I}(t)$,
it is clear that $a$ and $a^\dagger$ defined in
eq.~\eqref{eq:25} are hermitian conjugate to each other. Using
eq.~\eqref{eq:23} and eq.~\eqref{eq:25}, $\eta_I(t)$ and $p_{\eta I}(t)$
can be written, respectively, as
\begin{align}
  \eta_I(t)=a u(t)+a^{\dagger}v(t)\,,\qquad
p_{\eta I}(t)=a \dot{u}(t)+a^{\dagger}\dot{v}(t)\,.
\label{eq:9}
\end{align}
It should be noted that eq.~\eqref{eq:9} is valid not only for
real $t$ but also for imaginary $t$.

\subsection{In-in formalism along complex path}
For later convenience, we introduce the state $\ket{\Phi_N}$, which is
the state at the nucleation point when non-linear interactions are
switched off.
By taking the limit $t\to\pm i\infty$ in eq.~\eqref{eq:47}, eq.~\eqref{eq:59},
eq.~\eqref{eq:18} and eq.~\eqref{eq:34}, we define $\omega_F$,
$a_F$, $\ket{n_F}$, $H_{0_F}$ and $E_{n_F}$.
Using those asymptotic quantities, $\ket{\Phi_N}$ is obtained as
\begin{align}
 \ket{\Phi_{N}}&= \lim_{T\to\infty}e^{E_{0_F}T}U^{(0)}(0,iT)\ket{0_{F}}\,,
\label{eq:20}
\end{align}
where the normalization factor $e^{E_{0_F}T}$ is introduced to 
make the expression finite and constant in the limit $T\to\infty$.
As a result of the explicit $t$-dependence of
the free Hamiltonian $H_0(t)$,
$\ket{\Phi_{N}}$ is not proportional to $\ket{0_{F}}$ in general.
The difference between $\ket{\Phi_{N}}$ and $\ket{0_{F}}$
is determined by solving the 
EOMs for the positive and negative frequency functions
given in eq.~\eqref{eq:38}\footnote{The effect of the explicit $t$-dependence of $H_0(t)$ was determined by
directly solving the Schr\"odinger equation in \cite{Yamamoto:1993mp}.
For the correspondence between this work and \cite{Yamamoto:1993mp}, see App.~\ref{sec:post-freq-funct}.}.

As will be confirmed below, the annihilation operator $a$ that annihilates
$\ket{\Phi_{N}}$ is associated with $u(t)$ and $v(t)$ defined with the
boundary conditions
\begin{align}
  u(t)\stackrel{t\to- i\infty}{\to} e^{-i\omega_{F}t}\,,
\qquad v(t)\stackrel{t\to+ i\infty}{\to} e^{i\omega_{F}t}\,,
\label{eq:26}
\end{align}
up to constant factors determined by the KG normalization.
Note that $u(t)$ and $v(t)$ satisfy the conditions for positive and negative frequency functions
given by eq.~\eqref{eq:14} and eq.~\eqref{eq:23}.
The corresponding annihilation
operator is defined by substituting $v(t)$ given by eq.~\eqref{eq:26}
into eq.~\eqref{eq:25}, and can be rewritten as
\begin{align}
 a&=-\frac{i}{\hbar}
U^{(0)}(0,t)
\left(\eta\dot{v}(t)-p_{\eta}v(t)\right)U^{(0)}(t,0)\nnmb
&\propto \lim_{T\to \infty}e^{-\omega_F T}U^{(0)}(0,iT)a_F U^{(0)}(iT,0)\,.
\label{eq:30}
\end{align}
In deriving the first equality we used eq.~\eqref{eq:35}, 
eq.~\eqref{eq:25} and the $t$-independence of $a$, and in deriving the second we used eq.~\eqref{eq:59}
in the limit $t\to i\infty$ along with eq.~\eqref{eq:26}. 
Then, using eq.~\eqref{eq:20} and
eq.~\eqref{eq:30}, we can explicitly show that
\begin{align}
 a\ket{\Phi_N}&\propto \lim_{T\to \infty}e^{(E_{0_F}-\omega_F) T}
U^{(0)}(0,iT)a_F U^{(0)}(iT,0)U^{(0)}(0,iT)\ket{0_{F}}
=0\,,
\label{eq:37}
\end{align}
as we stated above.

Now we will move from the Schr\"odinger picture to the interaction
picture.  By inserting $U^{(0)}(t_1,t_2)U^{(0)}(t_2,t_1)=1$ into
eq.~\eqref{eq:16} many times, and using eq.~\eqref{eq:35} and
eq.~\eqref{eq:20}, we obtain
 \begin{align}
 &\bra{\Phi} U^{(0)}(-i\infty,0)U^{(0)}(0,-i\infty)
 U(-i\infty,0)U(0,t)U^{(0)}(t,0)U^{(0)}(0,t)\nnmb[-5pt]
\big<\mathcal{O}\big>_y\ 
=\ &\frac{\quad \qquad\times \mathcal{O}U^{(0)}(t,0)
 U^{(0)}(0,t) U(t,0)U(0,i\infty)
 U^{(0)}(i\infty,0)U^{(0)}(0,i\infty)\ket{\Phi}}
{\braket{\Phi}{
U^{(0)}(-i\infty,0)U^{(0)}(0,-i\infty)U(-i\infty,0) 
U(0,i\infty)U^{(0)}(i\infty,0)U^{(0)}(0,i\infty)}{\Phi}}\nnmb
=&\ \frac{\bra{\Phi_N}U_I(-i\infty,t)\mathcal{O}_I(t)
 U_I(t,i\infty)\ket{\Phi_N}}
{\braket{\Phi_N}{U_I(-i\infty,i\infty)}{\Phi_N}}\,,
\label{eq:22}
 \end{align}
where the overall factors appearing in both numerator and denominator
cancel each other.
To make the correspondence between this result and that of the 
conventional in-in formalism\cite{Weinberg:2005vy,Maldacena:2002vr}
clearer,  
we can rewrite eq.~\eqref{eq:22} as
\begin{align}
 \big<\mathcal{O}(t)\big>&
= \frac{\big<P\left(\mathcal{O}_I(t)\exp\left[-\frac{i}{\hbar}
\int_C
H_I(t')dt'\right]
\right)\big>^{(N)}}
{\big<P\left(\exp\left[-\frac{i}{\hbar}
\int_C
H_I(t')dt'\right]
\right)\big>^{(N)}}\,,
\label{eq:1}
\end{align}
where the time integration path $C$ is 
given by eq.~\eqref{eq:32},
$\big<\mathcal{O}(t)\big>\equiv\big<\mathcal{O}\big>_y$, and $\big<\
\big>^{(N)}$ is defined as $\big<\mathcal{O} \big>^{(N)}\equiv
\braket{\Phi_N}{\mathcal{O}}{\Phi_N}/\innr{\Phi_N}{\Phi_N}$.  We can
deform the integration path in the denominator from $C$ to $i\infty\to
-i\infty$ using $P\left(\exp\left[-\frac{i}{\hbar}\int_{0\to t\to 0}
H_I(t')dt'\right]\right)=1$.

Since the annihilation operator $a$ annihilates $\ket{\Phi_N}$,
Wick's theorem can be used to evaluate eq.~\eqref{eq:1} as usual.
The $N$-point correlation function $\big<P\big(\eta_I(t_1)\eta_I(t_2)\cdots \eta(t_N)\big)\big>^{(N)}$ 
vanishes when $N$ is odd,  but is given by
\begin{align}
\big<P\big(\eta_I(t_1)\eta_I(t_2)\cdots \eta(t_N)\big)\big>^{(N)} 
&=\sum_{{\rm set\ of\ pairs}}\quad \prod_{{\rm pairs}}\big<P\big(\eta_I(t_i)\eta_I(t_j)\big)\big>^{(N)}\,,
\label{eq:42}
\end{align}
when $N$ is even.
Here, the 2-point correlation function $\big<P\big(\eta_I(t_1)\eta_I(t_2)\big)\big>^{(N)}$
can be evaluated as
\begin{align}
\big<P\big(\eta_I(t_1)\eta_I(t_2)\big)\big>^{(N)}&=
\begin{cases}
 u(t_1)v(t_2)&{\rm when\ }t_1{\rm\ precedes\ } t_2 {\rm\ along\ }C\,,\\
 u(t_2)v(t_1)&{\rm when\ }t_2{\rm\ precedes\ } t_1 {\rm\ along\ }C\,,
\end{cases}
\label{eq:41}
\end{align}
where $u(t)$ and $v(t)$ are given by eq.~\eqref{eq:26}.

Before closing this section, let us summarize what we have found.
The expression given in eq.~\eqref{eq:1} is in the same form as the
conventional in-in formalism, which is often used in quantum field theory
calculations involving interactions\cite{Weinberg:2005vy,Maldacena:2002vr}.
However, the time integration path $C:i\infty\to0\to t\to0\to-i\infty$
is different from the usual case, where the time integration path is
$t_0\to t\to t_0$ when the initial state is given at an initial time
$t_0$ or $-\infty(1-i\epsilon)\to t\to -\infty(1+i\epsilon)$ when the
initial state is given in the past infinity.  In our case, the
quasi-ground-state is chosen as the initial state of the false vacuum,
and the corresponding time is given as $t=\pm i\infty$ using the
instanton $y(\tau)$ defined with Euclidean time $\tau=it$.  In
eq.~\eqref{eq:1}, the imaginary part of the integration path $C$
corresponds to the evolution inside the barrier, or during tunneling,
while the real part corresponds to the evolution outside the
barrier, or after tunneling. 

\section{Application to toy model}
\label{sec:appl-toy-model}

\subsection{Toy model}
For illustration purposes, we explicitly apply the formalism obtained above
to a simple toy model.
We assume that the instanton is given by
 \begin{align}
 y(\tau)\approx
 \begin{cases}
   y_F& \left(-\infty<\tau< -\tau_W\right)\,,\quad\left(\tau_W<\tau< \infty\right)\,,\\
 y_N& \left(-\tau_W<\tau< +\tau_W\right)\,,\\
>y_N &\left( 0<t(=-i\tau)<\infty\right)\,,
\label{eq:36}
 \end{cases}
 \end{align}
where $\tau_W$ ($0<\tau_W$) is the wall size of the thin-wall instanton,
and that the potential $V_\eta(y,\eta)$ is given by 
\begin{align}
 V_\eta(y,\eta)=\frac{\omega^2}{2}\eta^2+\tilde{\lambda}(y)\,\eta^3\,,
\label{eq:28}
\end{align}
where the $y$-dependent coupling constant $\tilde{\lambda}(y)$ is
assumed to be effective only inside the potential barrier (i.e. $y_F<y<y_N$).
By substituting eq.~\eqref{eq:36} and eq.~\eqref{eq:28}
into eq.~\eqref{eq:47}, $H(\tau)=H_0+H_{int}(\tau)$ can be written as
 \begin{align}
H_0=\frac{p^2}{2}+\frac{\omega^2}{2}\eta^2-\frac{\hbar\omega}{2}\,,\qquad
 H_{int}(\tau) &\approx\lambda \delta\left(\tau- \tau_W\right)\eta^3
+\lambda \delta\left(\tau+ \tau_W\right)\eta^3\,,
\label{eq:46}
 \end{align}
where $\delta(x)$ is Dirac's delta function and
$\lambda=\int_{-\tau_W-0}^{-\tau_W+0}\tilde{\lambda}(\bar{y}(\tau))d\tau$.
Here, the eigenenergy of the quasi-ground-state is given by $E_F=\hbar\omega/2$,
since $H_{int}(\tau)$ vanishes around the false vacuum and
the quasi-ground-state is the ground state for $H_0$.
In the following, we denote the
ground state and the annihilation operator associated with $H_0$ as 
$\ket{0}$ and $a$, respectively.
We will calculate $\big<\eta\big>_y$, or $\big<\eta(t)\big>$,
using both the the Schr\"odinger and interaction
picture expressions, given in eq.~\eqref{eq:16} and eq.~\eqref{eq:1}, respectively.
Although $\big<\eta(t)\big>=0$ in the free theory calculation,
we obtain $\big<\eta(t)\big>(t)\neq0$ as a result of 
the effect of non-linear interaction.

\subsection{Calculation in Schr\"odinger picture}
To evaluate eq.~\eqref{eq:16}, we obtain $\ket{\Phi(t)}$ using eq.~\eqref{eq:15}.
The evolution of the ground state $\ket{0}$ defined at the false vacuum ($t'=+i\infty$)
to behind the wall ($t'=-i(-\tau_W-0)$) is trivial
since $H(t')$ is simply given by $H_0$ in this region,
and we obtain
 \begin{align}
 \ket{\Phi(-i(-\tau_W-0))}&= \ket{0}\,.
\label{eq:49}
 \end{align}
Using eq.~\eqref{eq:46}, the evolution of the state across the wall
(i.e. $t'=-i(-\tau_W-0)\to -i(-\tau_W+0)$) is given by
 \begin{align}
 \ket{\Phi(-\tau_W+0)}&=e^{-\frac{\lambda}{\hbar} \eta^3}\ket{\Phi(-\tau_W-0)}\,.
\label{eq:48}
 \end{align}
Since $H(t')$ is again simply $H_0$ from in front of the wall ($t'=-i(-\tau_W+0)$)
to outside the barrier ($t'=t$),
the evolution of the state between them is given by
 \begin{align}
 \ket{\Phi(t)}&=\exp\left[-\frac{i}{\hbar}H_0
(t-i\tau_W)\right]\ket{\Phi(-i(-\tau_W+0))}\,.
\label{eq:50}
 \end{align}
By combining eq.~\eqref{eq:49}, eq.~\eqref{eq:48} and eq.~\eqref{eq:50}
we obtain, to first order in $\lambda$, 
\begin{align}
 \ket{\Phi(t)}&\approx\exp\left[-\frac{i}{\hbar}H_0
(t-i\tau_W)\right](1-\frac{\lambda}{\hbar}\eta^3)\ket{0}\,,
\label{eq:51}
\end{align}
and its hermitian conjugate is given by
\begin{align}
  \bra{\Phi(t)}&\approx\bra{0}(1-\frac{\lambda}{\hbar}\eta^3)
\exp\left[\frac{i}{\hbar}H_0
(t+i\tau_W)\right]\,.
\label{eq:52}
\end{align}
By substituting eq.~\eqref{eq:51} and eq.~\eqref{eq:52} into eq.~\eqref{eq:16}
we obtain, to leading order in $\lambda$, 
\begin{align}
 \big<\eta(t)\big>&\approx- \frac{\lambda}{\hbar}\braket{0}{
\eta \exp\left[-\frac{i}{\hbar}H_0
(t-i\tau_W)\right]\eta^3+
\eta^3 \exp\left[\frac{i}{\hbar}H_0
(t+i\tau_W)\right]\eta
}{0}\nnmb
&=-\frac{3\hbar\lambda}{2\omega^2}\cos\left(\omega t\right)e^{-\omega\tau_W}\,.
\label{eq:43}
\end{align}
To obtain the second line, we used $[a,a^\dagger]=1$, $H_0\ket{0}=0$, 
$[H_0,a]=-\hbar \omega$, $[H_0,a^\dagger]=\hbar \omega$
and $\eta=(\hbar/2\omega)^{1/2}\left(a+a^\dagger\right)$.

\subsection{Calculation in interaction picture}
Since $H_0$ is independent of $t$, eq.~\eqref{eq:38} can be easily solved.
$u(t)$ and $v(t)$ defined with the
boundary conditions in eq.~\eqref{eq:26}
are given, respectively, by
\begin{align}
  u(t)= \sqrt{\frac{\hbar}{2\omega}}e^{-i\omega t}\,,
\qquad v(t)= \sqrt{\frac{\hbar}{2\omega}}e^{i\omega t}\,.
\end{align}
By using $H_{int}(\tau)$ given in eq.~\eqref{eq:46} along with eq.~\eqref{eq:21},
we obtain, to first order in $\lambda$,
\begin{align}
 \exp\left[-\frac{i}{\hbar}
\int_C
H_I(t')dt'\right]&
\approx
1- \frac{\lambda}{\hbar}\eta_I^3(i\tau_W)
- \frac{\lambda}{\hbar}\eta_I^3(-i\tau_W)\,.
\label{eq:63}
\end{align}
By substituting eq.~\eqref{eq:63} into eq.~\eqref{eq:22}
we obtain, to leading order in $\lambda$,
\begin{align}
\big<\eta(t)\big>
&\approx-\frac{\lambda}{\hbar}
\big<\eta_I(t)\eta_I^3(i\tau_W)+\eta_I^3(-i\tau_W)\eta_I(t)\big>^{(N)}\nnmb
&=-\frac{3\hbar\lambda}{2\omega^2}\cos\left(\omega t\right)e^{-\omega\tau_W}\,.
\label{eq:44}
\end{align}
which is in agreement with eq.~\eqref{eq:43}, as it should be.
To obtain the second line,
we used Wick's theorem, as in eq.~\eqref{eq:42}.
Here, for example, $\big<\eta_I(t_1)\eta_I^3(t_2)\big>^{(N)}$ can be evaluated as
\begin{align}
\big<\eta_I(t_1)\eta_I^3(t_2)\big>^{(N)}
&=3\big<\eta_I(t_1)\eta_I(t_2)\big>^{(N)}\big<\eta_I^2(t_2)\big>^{(N)}
=3u(t_1)u(t_2)v^2(t_2)\,.
\end{align}

\section{Conclusion}
\label{sec:conclusion}
We have studied a 2-dimensional tunneling system, where the tunneling
sector $y$ is non-linearly coupled to an oscillator $\eta$.  Assuming
the system is initially in a quasi-ground-state at the false vacuum, the
2-dimensional tunneling wave function $\psi(y,\eta)$ has been
constructed using the WKB method. We have considered the effect of non-linear
interactions, which has not been studied in the context of multi-dimensional tunneling
systems before, to our knowledge.

We have determined the quantum expectation values with respect to the $\eta$
direction at a given $y$ outside the barrier.  We first introduced a
Schr\"odinger picture formulation to obtain eq.~\eqref{eq:16} in
Sec.~\ref{sec:wkbnext}, and then moved to an interaction picture formulation in
Sec.~\ref{sec:interaction} to obtain eq.~\eqref{eq:1}.  The resulting
formula given in eq.~\eqref{eq:1} is of the same form as the conventional in-in
formalism, which is often used in quantum field theory calculations with
interactions\cite{Weinberg:2005vy,Maldacena:2002vr}.  However, the time
integration path is modified to the one consisting of 
an imaginary part in addition to a real part.

The difference in the integration path
for the usual case and the quantum tunneling case  can be understood as follows.  In the usual case,
an initial state is given at some finite past $t=t_0$ or the infinite past
$t=-\infty$, both of which are defined on the real axis.  However, in
the case of quantum tunneling, the initial state is given at the
false vacuum, where the corresponding time is  $t=\pm i\infty$.
In our case, the imaginary part of the integration path corresponds to the
evolution of the quantum state during tunneling, while the real part
corresponds to the evolution after the quantum tunneling.

In this paper, the formulation has been done in a multi-dimensional quantum mechanical
system.  In order to apply it to cosmology, we need to
extend the formulation to field theory, with gravitational effects included.
Such an extension has been done in the case without interactions in
\cite{Yamamoto:1993mp,Tanaka:1993ez,Tanaka:1994qa}, and we expect
similar extension to be possible in the case with interactions. Although
a full derivation is now under investigation, one might naively expect
that the integration path will also consist of an imaginary part
corresponding to the evolution during quantum tunneling,
and real part corresponding to the evolution after quantum tunneling.
Calculations assuming this naive expectation to be true
have already been performed in
the literature\cite{Sugimura:2012kr,Park:2011ty}.

Observable effects resulting from non-linear interactions, such as the
non-Gaussianity of cosmological fluctuations, are now recognized as
powerful tools to probe the early universe.  It is therefore
important for us to be able to determine such features that may result
from models involving quantum tunneling, which are motivated by the string
landscape.

\begin{acknowledgments}
KS thanks J.~White, M.~Sasaki, T.~Tanaka and K.~Yamamoto for useful discussions and 
valuable comments.
This work was supported in part by Monbukagaku-sho 
Grant-in-Aid for the Global COE programs, 
``The Next Generation of Physics, Spun from Universality 
and Emergence'' at Kyoto University.
KS was supported by Grant-in-Aid for JSPS Fellows
No.~23-3437. 
\end{acknowledgments}

\appendix
\section{Positive frequency function and wave function}
\label{sec:post-freq-funct}
In this appendix, we will illustrate the the relation between the positive
frequency function $u(t)$ used in this work and its corresponding wave function
$\psi(\eta,t)$ used in the literature\cite{Yamamoto:1993mp}. 
We will employ the 1-dimensional harmonic oscillator with Hamiltonian
\begin{align}
 H&=\frac{p_\eta^2}{2}+\frac{\omega^2}{2}\eta^2\,,
\end{align}
as an example.
We will also see how the freedom in choosing $u(t)$ and $v(t)$ 
is related to the Bogolubov transformation.

As usual, the ground state $\ket{0}$ and the corresponding annihilation operator $a$ are given by
\begin{align}
a&=\sqrt{\frac{\omega}{2\hbar}}\eta+i\frac{1}{\sqrt{2\hbar\omega}}p_\eta\,,\qquad
a\ket{0}=0
\,,
\label{eq:54}
\end{align}
where the Hamiltonian can be rewritten as
$H=\hbar\omega\left(a^\dagger a +\frac{1}{2}\right)$
and the commutation relation is given by $\left[a,a^\dagger\right]=1$.
The Bogolubov transformed vacuum
state $\ket{\tilde{0}}$ and corresponding annihilation operator $\tilde{a}$ are constructed as
\begin{align}
\tilde{a}=\alpha a +\beta a^\dagger\,,\qquad
\tilde{a}\ket{\tilde{0}}&=0\,,
\label{eq:64}
\end{align}
where $\alpha$ and $\beta$ satisfy $|\alpha|^2-|\beta|^2=1$.
Here, $\tilde{a}$ satisfies the commutation relation 
$\left[\tilde{a},\tilde{a}^\dagger\right]=1$ but is nothing to do with the Hamiltonian.

In the Heisenberg picture, operators are defined as $\mathcal{O}_H(t)
=e^{\frac{i}{\hbar}Ht}\mathcal{O} e^{-\frac{i}{\hbar}Ht}$, where
operators with and without subscript $H$ correspond to Heisenberg and
Schr\"odinger operators, respectively.  
The positive frequency functions
$u(t)$, which are solutions to the EOM $\ddot{u}(t)=-\omega^2 u(t)$ and
satisfy the Klein-Gordon normalization $u\dot{u}^*-\dot{u}u^*=i\hbar$,
define the corresponding annihilation operators $a_u$ by
\begin{align}
 a_u&=\frac{1}{i}\left(\eta_H(t)\dot{u}^*(t)-p_{\eta H}(t)u^{*}(t)\right)\,.
\label{eq:53}
\end{align}
The positive frequency function $u_0(t)=\sqrt{\hbar/2\omega}\,e^{-i\omega t}$
gives the annihilation operator $a$ of the ground state defined in eq.~\eqref{eq:54},
while $\tilde{u}(t)=\alpha^* u_0(t)-\beta u_0^*(t)$ gives $\tilde{a}$ of the Bogolubov transformed vacuum state
defined in eq.~\eqref{eq:64}.

We will explicitly construct the wave function
$\psi_u(\eta)=\innr{\eta}{0_u}$ where $\ket{0_u}$ satisfies $a_u\ket{0_u}=0$. Using
eq.~\eqref{eq:53}, $\mathcal{O}_H(t) =e^{\frac{i}{\hbar}Ht}\mathcal{O}
e^{-\frac{i}{\hbar}Ht}$ and $p_\eta=-i\hbar (\partial/\partial\eta)$,
we can rewrite $a_u\ket{0_u}=0$ in terms of the wave function as
\begin{align}
\left(i\hbar\, u^{*}(t)\frac{\partial}{\partial \eta}+\dot{u}^*(t)\eta\right)
e^{-\frac{i}{\hbar}Ht}\psi_u(\eta,t)&=0\,,
\label{eq:57}
\end{align}
where $H=-(\hbar^2/2)(\partial^2/\partial \eta^2) +(\omega^2/2)\eta^2$.

For the ground state, the positive frequency function is given by $u_0(t)$
and $H=\hbar\omega/2$. By solving eq.~\eqref{eq:57}, we obtain, neglecting an imaginary phase,
\begin{align}
\psi_0(\eta)=\sqrt{\frac{\omega}{\pi\hbar}}\exp\left[-\frac{\omega \eta^2}{2\hbar}\right]\,,
\label{eq:31}
\end{align}
where $\psi_0(\eta)$ is the well known ground state wave function for the harmonic oscillator,
as expected.
Here we choose the overall normalization such that $\int d\eta |\psi_0(\eta,t)|^2=1$.

For the Bogolubov transformed vacuum state, 
the positive frequency function is given by $\tilde{u}(t)$.
Using the hermiticity of $H$ and solving eq.~\eqref{eq:57}, we obtain, neglecting an imaginary phase,
\begin{align}
\tilde{\psi}(\eta,t)=e^{\frac{i}{\hbar}Ht}
\left(\sqrt{\frac{1}{\pi\hbar }\frac{\dot{\tilde{u}}^*(t)}{\tilde{u}^{*}(t)}}
\exp\left[\frac{i}{2\hbar}\frac{\dot{\tilde{u}}^*(t)}{\tilde{u}^{*}(t)} \eta^2\right]\right)\,,
\label{eq:55}
\end{align}
where we again choose the overall normalization such that $\int d\eta |\tilde{\psi}(\eta,t)|^2=1$.

\bibliography{$HOME/physics/mybib}

\providecommand{\href}[2]{#2}\begingroup\raggedright\begin{thebibliography}{10}

\bibitem{mohsen:2003}
M.~Razavy, {\em Quantum Theory of Tunneling}.
\newblock World Scientific Pub Co Inc, 2003.

\bibitem{coleman_aspects}
S.~Coleman, {\em Aspects of Symmetry: Selected Erice Lectures}.
\newblock Cambridge University Press, 1988.

\bibitem{Vilenkin:1984wp}
A.~Vilenkin, {\it {Quantum Creation of Universes}},  {\em Phys. Rev.} {\bf D30}
  (1984) 509--511.

\bibitem{Susskind:2003kw}
L.~Susskind, {\it {The anthropic landscape of string theory}},  {\em
  arXiv:hep-th/0302219} (2003)
  [\href{http://xxx.lanl.gov/abs/hep-th/0302219}{{\tt hep-th/0302219}}].

\bibitem{Coleman:1977py}
S.~R. Coleman, {\it {The Fate of the False Vacuum. 1. Semiclassical Theory}},
  {\em Phys. Rev.} {\bf D15} (1977) 2929--2936. [Erratum-ibid.D16:1248,1977].

\bibitem{Callan:1977pt}
J.~Callan, Curtis~G. and S.~R. Coleman, {\it {The Fate of the False Vacuum. 2.
  First Quantum Corrections}},  {\em Phys.Rev.} {\bf D16} (1977) 1762--1768.

\bibitem{Coleman:1980aw}
S.~R. Coleman and F.~De~Luccia, {\it {Gravitational Effects on and of Vacuum
  Decay}},  {\em Phys. Rev.} {\bf D21} (1980) 3305.

\bibitem{Sugimura:2011tk}
K.~Sugimura, D.~Yamauchi, and M.~Sasaki, {\it {Multi-field open inflation model
  and multi-field dynamics in tunneling}},  {\em JCAP} {\bf 1201} (2012) 027,
  [\href{http://xxx.lanl.gov/abs/1110.4773}{{\tt arXiv:1110.4773}}].

\bibitem{Banks:1973ps}
T.~Banks, C.~M. Bender, and T.~T. Wu, {\it {Coupled anharmonic oscillators. 1.
  Equal mass case}},  {\em Phys.Rev.} {\bf D8} (1973) 3346--3378.

\bibitem{Banks:1974ij}
T.~Banks and C.~M. Bender, {\it {Coupled anharmonic oscillators. ii.
  unequal-mass case}},  {\em Phys.Rev.} {\bf D8} (1973) 3366--3378.

\bibitem{Gervais:1977nv}
J.-L. Gervais and B.~Sakita, {\it {WKB Wave Function for Systems with Many
  Degrees of Freedom: A Unified View of Solitons and Instantons}},  {\em
  Phys.Rev.} {\bf D16} (1977) 3507.

\bibitem{Yamamoto:1993mp}
K.~Yamamoto, {\it {Quantum tunneling in multidimensional systems}},  {\em Prog.
  Theor. Phys.} {\bf 91} (1994) 437--452.

\bibitem{Tanaka:1993ez}
T.~Tanaka, M.~Sasaki, and K.~Yamamoto, {\it {Field theoretic description of
  quantum fluctuations in multidimensional tunneling approach}},  {\em Phys.
  Rev.} {\bf D49} (1994) 1039--1046.

\bibitem{Tanaka:1994qa}
T.~Tanaka and M.~Sasaki, {\it {Quantum state during and after O(4) symmetric
  bubble nucleation with gravitational effects}},  {\em Phys. Rev.} {\bf D50}
  (1994) 6444--6456, [\href{http://xxx.lanl.gov/abs/gr-qc/9406020}{{\tt
  gr-qc/9406020}}].

\bibitem{Yamamoto:1996qq}
K.~Yamamoto, M.~Sasaki, and T.~Tanaka, {\it {Quantum fluctuations and CMB
  anisotropies in one-bubble open inflation models}},  {\em Phys. Rev.} {\bf
  D54} (1996) 5031--5048, [\href{http://xxx.lanl.gov/abs/astro-ph/9605103}{{\tt
  astro-ph/9605103}}].

\bibitem{Garriga:1997wz}
J.~Garriga, X.~Montes, M.~Sasaki, and T.~Tanaka, {\it {Canonical quantization
  of cosmological perturbations in the one-bubble open universe}},  {\em Nucl.
  Phys.} {\bf B513} (1998) 343--374,
  [\href{http://xxx.lanl.gov/abs/astro-ph/9706229}{{\tt astro-ph/9706229}}].

\bibitem{Garriga:1998he}
J.~Garriga, X.~Montes, M.~Sasaki, and T.~Tanaka, {\it {Spectrum of cosmological
  perturbations in the one-bubble open universe}},  {\em Nucl. Phys.} {\bf
  B551} (1999) 317--373, [\href{http://xxx.lanl.gov/abs/astro-ph/9811257}{{\tt
  astro-ph/9811257}}].

\bibitem{Bartolo:2004if}
N.~Bartolo, E.~Komatsu, S.~Matarrese, and A.~Riotto, {\it {Non-Gaussianity from
  inflation: Theory and observations}},  {\em Phys.Rept.} {\bf 402} (2004)
  103--266, [\href{http://xxx.lanl.gov/abs/astro-ph/0406398}{{\tt
  astro-ph/0406398}}].

\bibitem{Komatsu:2009kd}
E.~Komatsu, N.~Afshordi, N.~Bartolo, D.~Baumann, J.~Bond, {\em et.~al.}, {\it
  {Non-Gaussianity as a Probe of the Physics of the Primordial Universe and the
  Astrophysics of the Low Redshift Universe}},
  \href{http://xxx.lanl.gov/abs/0902.4759}{{\tt arXiv:0902.4759}}.

\bibitem{Chen:2010xka}
X.~Chen, {\it {Primordial Non-Gaussianities from Inflation Models}},  {\em
  Adv.Astron.} {\bf 2010} (2010) 638979,
  [\href{http://xxx.lanl.gov/abs/1002.1416}{{\tt arXiv:1002.1416}}].

\bibitem{Sugimura:2012kr}
K.~Sugimura, D.~Yamauchi, and M.~Sasaki, {\it {Non-Gaussian bubbles in the
  sky}},  {\em Europhys.Lett.} {\bf 100} (2012) 29004,
  [\href{http://xxx.lanl.gov/abs/1208.3937}{{\tt arXiv:1208.3937}}].

\bibitem{Park:2011ty}
D.~S. Park, {\it {Scalar Three-point Functions in a CDL Background}},  {\em
  JHEP} {\bf 1201} (2012) 165, [\href{http://xxx.lanl.gov/abs/1111.2858}{{\tt
  arXiv:1111.2858}}].

\bibitem{Weinberg:2005vy}
S.~Weinberg, {\it {Quantum contributions to cosmological correlations}},  {\em
  Phys. Rev.} {\bf D72} (2005) 043514,
  [\href{http://xxx.lanl.gov/abs/hep-th/0506236}{{\tt hep-th/0506236}}].

\bibitem{Maldacena:2002vr}
J.~M. Maldacena, {\it {Non-Gaussian features of primordial fluctuations in
  single field inflationary models}},  {\em JHEP} {\bf 0305} (2003) 013,
  [\href{http://xxx.lanl.gov/abs/astro-ph/0210603}{{\tt astro-ph/0210603}}].

\end{thebibliography}\endgroup

\end{document}